\documentclass
[article,superscriptaddress,tightenlines,amsmath,amssymb]
{revtex4}

\usepackage{amssymb}
\usepackage{graphicx}
\usepackage{subfigure}
\usepackage{color}
\usepackage{hyperref}
\hypersetup{
 colorlinks,
linkcolor=red,
filecolor=green,
urlcolor=blue,
citecolor=blue
}

\newcommand{\bea}{\begin{eqnarray}}
\newcommand{\eea}{\end{eqnarray}}

\newcommand{\mrbeta}{{{}_{0}\beta}}

\newcommand{\ket}[1]{\left| {#1} \right\rangle}

\begin{document}

\title{Towards universal quantum computation through relativistic motion}

\author{David Edward Bruschi}
\address{School of Mathematical Sciences, University of Nottingham, University Park,
Nottingham NG7 2RD, United Kingdom}
\address{School of Electronic and Electrical Engineering, University of Leeds, Woodhouse Lane,  Leeds, LS2 9JT,  United Kingdom. \footnote{York Centre for Quantum Technologies, Department of Physics, University of York, Heslington,  YO10 5DD York, UK}}
\author{Carlos Sab{\'\i}n}
\address{Instituto de F{\'\i}sica Fundamental, CSIC, Serrano 113-bis,
28006 Madrid, Spain\\ csl@iff.csic.es}
\author{Pieter Kok}
 \address{Department of Physics \& Astronomy, University of Sheffield, Sheffield S3 7RH, United Kingdom}
 \author{G\"oran Johansson}
 \address{Microtechnology and Nanoscience, MC2, Chalmers University of Technology, S-41296 G\"oteborg, Sweden}
 \author{Per Delsing}
 \address{Microtechnology and Nanoscience, MC2, Chalmers University of Technology, S-41296 G\"oteborg, Sweden}
 \author{Ivette Fuentes}
 \address{School of Mathematical Sciences, University of Nottingham, University Park,
Nottingham NG7 2RD, United Kingdom}
\newpage

\begin{abstract}
We show how to use relativistic motion to generate continuous variable Gaussian cluster states within cavity modes. Our results can be demonstrated experimentally using superconducting circuits where tuneable boundary conditions correspond to mirrors moving with velocities close to the speed of light. In particular, we propose the generation of a quadripartite square cluster state as a first example that can be readily implemented in the laboratory. Since cluster states are universal resources for universal one-way quantum computation, our results pave the way for relativistic quantum computation schemes.
\end{abstract}
\maketitle



\begin{figure}[h]
\includegraphics[width=\columnwidth]{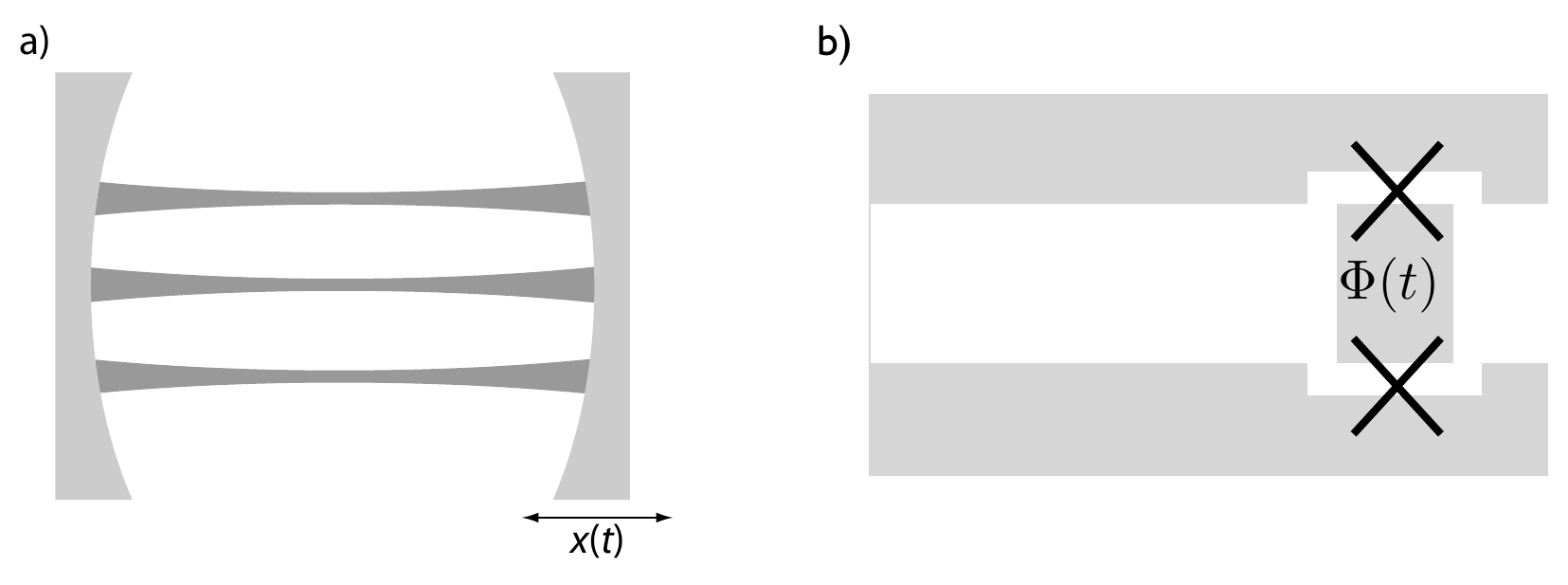}
\caption{Schematic of the physical setup. a) Cluster states on optical modes in a cavity with a moving mirror. The dynamical Casimir effect due to the accelerated motion of the mirror generates two-mode squeezing in the cavity modes. b) Creating cluster states in a superconducting circuit: the time-varying current changes the flux through the SQUID, leading to a dynamical Casimir effect equivalent to that in a).}
\label{fig:cavity}
\end{figure}

Quantum technologies are expected to bring great benefits to many human endeavours in the not-too-distant future. In the past decades it was shown that devices exploiting the laws of quantum mechanics can in principle cross boundaries that classical devices cannot. Advanced systems such as quantum memories and quantum computers are being developed in an effort to realise these transformative quantum technologies \cite{reviewnature}. In addition to the practical aspects of quantum technologies, there is great interest in viewing quantum information processing as a fundamental description of Nature. This leads naturally to the question how quantum information processing is affected by relativistic motion \cite{ivyalsingreview}. Here, we ask the question how the relativistic motion of a cavity mirror can be used to achieve a quantum state inside the cavity that can in principle be used for quantum computation. We show that we can create such a universal resource state in the continuous variables defined by the quadratures of the field modes inside the cavity. This proposal can be implemented practically using superconducting circuits (see Fig.~\ref{fig:cavity}).

Recently, it was shown that relativistic motion can be used to generate bipartite quantum gates. In particular, two-mode squeezing \cite{relgates} and beam-splitting gates \cite{jormadaniele} between field modes were implemented through the non- uniform acceleration of a cavity. A promising experimental realisation of these scenarios is circuit Quantum Electrodynamics (cQED), where Superconducting Quantum Interferometric Devices (SQUIDs) provide tuneable boundary conditions corresponding to mirrors moving at speeds close to the speed of light in the medium. This possibility was exploited for the first experimental observation of the Dynamical Casimir Effect, i.e., particle creation through relativistic motion of boundary conditions \cite{moore,casimirwilson}. We note that the beam-splitting gate was recently experimentally demonstrated in this type of system \cite{aumentado} and that two-mode squeezing involving two distinct cavity modes has also been achieved in cQED \cite{devoret}.Moreover, some of us have shown that the same phenomenon originates an observable effect in the fidelity of quantum teleportation between two superconducting cavities, one of them undergoing a temporal variation of the boundary conditions \cite{teleportationico}. These works open a new avenue of research on the effects of relativistic motion and gravity in generic quantum technologies, and in particular quantum computing with continuous quantum variables. While it is possible to transfer the continuous-variables entanglement generated in the Dynamical Casimir Effect to discrete-variables systems such as qubits \cite{felicetti} the analysis of relativistic effects in qubits is typically limited by the impossibility of achieving relativistic velocities -which can only be accessed in a a quantum simulator \cite{felicetti2}. 

Quantum computation with continuous variables (CV) has attracted a great deal of attention as an alternative to the traditional qubit approach to quantum computing \cite{reviewgaussian}, and the CV version of measurement-based quantum computing is well-developed \cite{braunstein,minion1, minion2,minion3}. In this paradigm for quantum computing the quantum gates are implemented by measurements on a particular type of multipartite entangled state known as a \emph{cluster state}. Assuming that we can implement the required measurements (via homodyne detection and photon counting \cite{GKP01}), these states are universal for quantum computation. A practical way to create CV cluster states in a cavity using non-relativistic methods is by creating an entangled optical frequency comb \cite{menicucci08}, a proof-of-principle experiment of which was recently performed \cite{pfister}. 

In this paper we show how to generate large CV regular cluster states through the relativistic accelerated motion of a cavity. The quantum field inside the cavity is initially in the vacuum state. We exploit entanglement resonances among different modes of the cavity \cite{relgates} to show that, by suitably choosing the motion of the boundaries, it is possible to generate specific multimode entangled states. These states can always be transformed into regular CV cluster states by suitably phase-shifting some of the modes. We propose an specific implementation with superconducting cavities in which the motion of the boundary conditions is implemented through tunable SQUIDs. In particular, we show an example with four modes that is within reach of current technology. In this way, we also contribute to transferring the concept of measurement based quantum information processing using continuous variables to the field of cQED.

In the Methods we will review the technical tools required to describe quantum fields in relativistically moving cavities. In the Results section we show how to produce CV cluster states using relativistic motion and we present an experimental implementation in superconducting circuits. We conclude our paper with a discussion of our results.

\section*{Methods}
\subsection*{Technical tools: quantum fields confined within moving cavities}\noindent
In this section we review the technical tools to describe relativistic quantum fields, following mainly \cite{relgates} unless stated otherwise. We consider that the length of the cavity is constant in the reference frame of an observer co-moving with the cavity. Using the covariance matrix formalism, we show how to use entanglement resonances to create a two-mode squeezing gate in the case the motion is either discrete or continuous. Finally, we show that the gates can also be produced in case a single cavity wall oscillates. 

\subsubsection*{Field quantization within a cavity}
We consider a massless field $\Phi$ confined within a $1+1$ cavity in flat space-time, which can be employed to model a single polarisation-mode of light \cite{louko:lee}. We use Minkowski coordinates $(t,x)$ which are a convenient choice to describe the field when the cavity undergoes inertial motion (in this work the signature is $(-,+)$). The left and right boundaries of the cavity are initially at positions $x_L$ and $x_R$ respectively, where $x_L<x_R$, the length of the cavity is $L:=x_R-x_L$ and we impose Dirichlet boundary conditions at the walls of the cavity, i.e. $\Phi(t,x_L)=\Phi(t,x_R)=0$. When the cavity is inertial, the mode solutions $\phi_n(t,x)$ to the Klein-Gordon equation $\square \Phi=0$ form a discrete set and take the form
\begin{align}
\phi_n(t,x)=\frac{1}{\sqrt{2\pi\,k_n}}e^{-it\omega_n}\sin(k_n (x-x_L)),
\end{align}
where the Minkowski frequencies are defined as $\omega_n=c\,k_n=\pi\,n\,c/L$ and the modes $\phi_n(t,x)$ satisfy the eigenvalue equation
\begin{align}
i\partial_t\phi_n(t,x)=\omega_n \phi_n(t,x).
\end{align}
The field $\Phi$ can be expanded as 
\begin{align}
\Phi=\sum_{n=0}^{+\infty}\left[\phi_n(t,x)a_n+\phi_n(t,x)^*a^{\dagger}_n\right],
\end{align}
where the operators $\{a_n,a^{\dag}_n\}$ are bosonic operators which satisfy the canonical commutation relations $[a_n,a^{\dagger}_m]=\delta_{nm}$ and define the vacuum by $a_n \ket{0}=0$ for all $n\in\mathbb{N}$.

We consider that at $t=0$ the cavity undergoes a period of uniform acceleration. Our formalism can be used to describe both discrete and continuous changes in acceleration.  Rindler coordinates  $(\eta,\chi)$  are a suitable choice to describe the cavity during periods of accelerated motion. The transformtation between Minkowski and Rindler coordinates is given by
\begin{align}
t=&\frac{\chi}{c}\sinh (\eta) \nonumber\\
x=&\chi\cosh\big(\eta),
\end{align}
where $\chi>0$ has dimensions of length and $\eta\in\mathbb{R}$ is the dimensionless Rindler time coordinate.
The proper acceleration of an observer moving along a trajectory $\chi=\sqrt{c^2t^2-x^2}=$const is $a=c^2/\chi$. A crucial point is that we consider that the cavity length is constant with respect to this observer. The observer's proper time $\tau$ is related to the Rindler time coordinate by $\tau={\chi}\eta/c$. Thus we can write $t=\frac{c}{a} \sinh{\frac{a\tau}{c}}$, $x=\frac{c^2}{a} \cosh{\frac{a\tau}{c}}$ and  $\tau=\frac{c}{a}\operatorname{arctanh}\big(\frac{c\,t}{x}\big)$, as has been done in \cite{relgates}.The solutions $\tilde{\phi}_n$ to the Klein-Gordon equation in Rindler coordinates take the form
\begin{align}
\tilde{\phi}_n(\tau,\chi)=\frac{1}{\sqrt{2\pi K_n}}e^{-i\tau\,\Omega_n}\sin\left[K_n\left(\chi-\chi_L\right)\right],
\end{align}
where the Rindler frequencies are defined as $\Omega_n=c\,K_n=n\,\pi\,c/(\chi_R-\chi_L)$.   The Rindler modes $\tilde{\phi}_n$ satisfy the eigenvalue equation
\begin{align}
i\partial_{\tau}\tilde{\phi}_n(\tau,\chi)=\Omega_n \tilde{\phi}_n(\tau,\chi).
\end{align}
During the periods of uniform acceleration, the field $\Phi$ can be therefore expanded in these coordinates as
\begin{align}
\Phi=\sum_{n=0}^{+\infty}\left[\tilde{\phi}_n(\tau,\chi)\tilde{a}_n+\tilde{\phi}_n(\tau,\chi)^*\tilde{a}^{\dagger}_n\right],
\end{align}
where the annihilation operators $\tilde{a}_n$ define the Rindler vacuum $|\tilde{0}\rangle$ by $\tilde{a}_n\,|\tilde{0}\rangle=0$ for all $n\in\mathbb{N}$ and satisfy the canonical commutation relations $[\tilde{a}_n,\tilde{a}_n^{\dagger}]=\delta_{nm}$.

\subsubsection*{Bogoliubov  transformation}
It is well known that the vacua $|\tilde{0}\rangle$ and $|0\rangle$ are different \cite{birrelldavies}. This implies that an accelerated observer will disagree with an inertial one that the state $|0\rangle$ is devoid of particles. This fundamental observation is at the basis of  the most exciting phenomena predicted by quantum field theory in curved spacetimes---such as dynamical Casimir effect and Unruh-Hawking effect---and is also responsible for relativistic effects in Quantum Information tasks between moving observers \cite{teleportationico}.

We now briefly review the techniques introduced in Refs.~\cite{alphacentauri,nicoivy} to study quantum information in cavities undergoing relativistic motion.
The mode solutions to the Klein-Gordon equation in Minkowski and Rindler coordinates are related by the following Bogoliubov transformations \cite{birrelldavies}, 
\begin{align}
\begin{pmatrix}
\tilde{\phi}\\
\tilde{\phi}^{\dagger}
\end{pmatrix}
=
\begin{pmatrix}
{}_0\boldsymbol{\alpha} & {}_0\boldsymbol{\beta}\\
{}_0\boldsymbol{\beta}^* & {}_0\boldsymbol{\alpha}^*
\end{pmatrix}
\begin{pmatrix}
\phi\\
\phi^{\dagger}
\end{pmatrix},\label{mode:bogo:transformation}
\end{align}
where the coefficients of the infinite-dimensional matrices ${}_0\boldsymbol{\alpha}$ and ${}_0\boldsymbol{\beta}$ are defined by
\begin{eqnarray}
{}_0\alpha_{mn}=&\left.(\tilde{\phi}_m,\phi_n)\right|_{t=0}\nonumber\\
{}_0\beta_{mn}=&\left.(\tilde{\phi}_m,\phi^{\star}_n)\right|_{t=0}\label{bogo:coefficient:definition}
\end{eqnarray}
and $(\cdot,\cdot)$ is the conserved inner product, see \cite{birrelldavies}.
Therefore, the induced transformation between the bosonic operators is
\begin{align}
\begin{pmatrix}
\tilde{a}\\
\tilde{a}^{\dagger}
\end{pmatrix}
=
\begin{pmatrix}
{}_0\boldsymbol{\alpha}^{\dagger} & -{}_0\boldsymbol{\beta}^T\\
-{}_0\boldsymbol{\beta}^{\dagger} & {}_0\boldsymbol{\alpha}^T 
\end{pmatrix}
\begin{pmatrix}
a\\
a^{\dagger}
\end{pmatrix}.\label{bogo:transformation}
\end{align}
The Bogoliubov coefficients ${}_0\beta_{mn}$ account for particle creation, while ${}_0\alpha_{mn}$ account for mode mixing. It is possible to compute analytically the inner products in Eq. \eqref{bogo:coefficient:definition} by expanding the Bogoliubov coefficients in terms of a dimensionless parameter $h={a L}/{c^2}\ll1$, as was done in \cite{alphacentauri}, and one obtains
\begin{eqnarray}
{}_0\alpha_{mn}&=&{}_0\alpha_{mn}^{(0)}+{}_0\alpha_{mn}^{(1)}+{}_0\alpha^{(2)}_{mn}+\mathcal{O}(h^3)\nonumber\\
{}_0\beta_{mn}&=&{}_0\beta_{mn}^{(1)}+{}_0\beta_{mn}^{(2)}+\mathcal{O}(h^3),\label{bogo:expansion}
\end{eqnarray}
where the superscripts stands for the corresponding order in the perturbative expansion. Notice that the zero order coefficients ${}_0\alpha_{mn}^{(0)}$ are diagonal and can be written as ${}_0\alpha_{mn}^{(0)}\equiv\,G_n(t)\delta_{nm}$.

The explicit expression of the first contributions to \eqref{bogo:expansion} can be found in \cite{alphacentauri}. The Bogoliubov coefficients ${}_0\alpha_{mn}$ and ${}_0\beta_{mn}$ correspond to instantaneous changes between inertial and uniformly accelerated motion.  A general cavity trajectory can be described as a succession of such transformations followed by periods of free evolution during which the field modes acquire phases $G_n(t)=\exp(-i\omega_n t)$ and $\tilde{G}_n(\tau)=\exp(-i\Omega_n\tau)$ for inertial and uniformly accelerated free evolution respectively and \cite{alphacentauri,nicoivy}. Notice that $\tilde{G}_n=G_n+\mathcal{O}(h^3)$. The modes of a cavity at rest and the modes of the cavity after any trajectory are related by a general Bogoliubov transformation of the form \eqref{mode:bogo:transformation}, where the transformation matrix is  
\begin{align}
\begin{pmatrix}
\boldsymbol{A}^{\dagger} & -\boldsymbol{B}^T\\
-\boldsymbol{B}^{\dagger}  & \boldsymbol{A}^T 
\end{pmatrix},
\end{align}
the Bogoliubov matrices $\boldsymbol{A}$ and $\boldsymbol{B}$ are functions of the Minkowski-Rindler instantaneous Bogoliubov coefficients ${}_0\alpha_{mn},{}_0\beta_{mn}$ and the phases $G_n$ acquired during the periods of free evolution. The exact expression depends on the travel scenario chosen \cite{alphacentauri}.

\subsubsection*{Covariance Matrix formalism}
We use the covariance matrix formalism which is particularly convenient when one needs to apply quantum information techniques to quantum field theory~\cite{nicoivy}. In this formalism, which is applicable to bosonic fields in Gaussian states, all the relevant information about the state is encoded in the first and second moments of the field. In particular, given a collection of N bosonic modes, we can collect the second moments in the Covariance Matrix $\boldsymbol{\sigma}$ defined by the elements $\sigma_{ij}=\langle X_{i} X_{j}+X_{j}X_{i}\rangle-2\langle X_{i}\rangle\langle X_{j}\rangle$, where $\langle\,.\,\rangle$ denotes the expectation value with respect to the state of the field and the quadrature operators $X_{i}$ are the generalized position and momentum operators of the field modes. In this paper we follow the conventions used in Ref.~\cite{nicoivy}, where the operators for the $n$-th mode are given by 
\begin{eqnarray}\label{xp}
X_{2n-1} =&\,\frac{1}{\sqrt{2}}(a_{n}+a^{\dag}_{n}) \equiv& Q_n,\nonumber\\ 
X_{2n}=&\,\frac{-i}{\sqrt{2}}(a_{n}-a^{\dag}_{n}) \equiv& P_n,
\end{eqnarray}
and $[Q_n,P_m]=i\,\delta_{nm}$.

The covariance matrix formalism enables elegant and simplified calculations and has been proven useful to define and analyze computable measures of bipartite and multipartite entanglement for Gaussian states~\cite{reviewgaussian}.

Every unitary transformation $U$ in Hilbert space that is generated by a quadratic Hamiltonian can be represented as a symplectic matrix $\boldsymbol{S}$ in phase space. These transformations form the real symplectic group $Sp(2N,\mathbb{R})
$, the group of real $(2N\times2N)$ matrices that leave the symplectic form $\boldsymbol{\Omega}$ invariant, i.e., $\boldsymbol{S}\,\boldsymbol{\Omega}\,\boldsymbol{S}^{T}=\boldsymbol{\Omega}$, where the symplectic form takes the expression $\boldsymbol{\Omega}=\bigoplus_{i=1}^{n}\boldsymbol{\Omega}_i$ and 
$$\boldsymbol{\Omega}_i=\left(
            \begin{array}{cc}
              0 & 1 \\
              -1 & 0 \\
            \end{array}
          \right) .
$$ 
The time evolution of the field, as well as the Bogoliubov transformations, can be encoded in this structure. The symplectic matrix corresponding to the Bogoliubov transformation in Eq.~(\ref{mode:bogo:transformation}) can be written in terms of the Bogoliubov coefficients as
\begin{equation}\label{Bogosymplectic}
\boldsymbol{S}=\left(
  \begin{array}{cccc}
    \boldsymbol{\mathcal{M}}_{11} & \boldsymbol{\mathcal{M}}_{12} & \boldsymbol{\mathcal{M}}_{13} & \cdots \\
    \boldsymbol{\mathcal{M}}_{21} & \boldsymbol{\mathcal{M}}_{22} & \boldsymbol{\mathcal{M}}_{23} & \cdots \\
    \boldsymbol{\mathcal{M}}_{31} & \boldsymbol{\mathcal{M}}_{32} & \boldsymbol{\mathcal{M}}_{33} & \cdots \\
    \vdots & \vdots & \vdots & \ddots
  \end{array}
\right)\,,
\end{equation}
where the $\boldsymbol{\mathcal{M}}_{mn}$ are the $2\times2$ matrices which have the explicit form
\begin{equation}\label{Mmatrices}
\boldsymbol{\mathcal{M}}_{mn}=\left(
                   \begin{array}{cc}
                     \mathrm{Re}(A_{mn}-B_{mn}) & \mathrm{Im}(A_{mn}+B_{mn}) \\
                     -\mathrm{Im}(A_{mn}-B_{mn}) & \mathrm{Re}(A_{mn}+B_{mn})
                   \end{array}
                 \right)\,.
\end{equation}
The covariance matrix $\boldsymbol{\tilde{\sigma}}$ after a Bogoliubov transformation is given by $\boldsymbol{\tilde{\sigma}}=\boldsymbol{S}\,\boldsymbol{\sigma}\, \boldsymbol{S}^{T}$. Let us assume that we are only interested in the state of two modes~$k$ and~$k^{\prime}$ after the transformation. A great advantage of the covariance matrix formalism is that the trace operation over a mode is implemented simply by deleting the row and column corresponding to that mode. In the next subsection, we will apply this formalism to compute the transformed state of an initial vacuum of the field when the cavity undergoes motion of the boundary conditions.


\subsubsection*{Entanglement resonances and two-mode squeezer gates}
When a cavity undergoes nonuniform motion, entanglement  is created between every pair of modes \cite{relgates}. Furthermore, the entanglement between chosen couples of modes can be selectively enhanced by ``resonances''. An entanglement resonance is a linear increase of entanglement proportional to the number of times a particular travel scenario is repeated \cite{alphacentauri,relgates,multip,jormadaniele}.

There are two different kind of entanglement resonances in the system.  The first is connected to particle creation and leads to a two-mode squeezing gate  \cite{relgates}.  Particle creation resonances have been extensively studied in the context of the dynamical Casmir effect \cite{casimirwilson}. The second entanglement resonance is produced by mode mixing without particle creation, which implements beam splitting gates\cite{jormadaniele}. In this work we will exploit particle creation resonances, since the squeezing gates which can be produced by them are useful resource for the generation of cluster states \cite{minion3}. 

Let us  consider that the field is initially in the vacuum state. The covariance matrix in this case is the identity matrix in $2M$ dimensions where $M$ is the number of modes. The cavity then undergoes $N$ identical intervals of uniform acceleration, each of them of duration 
\begin{align}
T=\frac{2\pi}{\omega_k+\omega_{k'}}.
\end{align}
Here $\omega_n$ is the the frequency of mode $n$. The $4\times 4$ reduced covariance matrix of two oddly-separated modes $k$ and $k'$ is obtained by tracing over Eq. (\ref{Bogosymplectic}) yielding, 
\begin{equation}\label{eq:sigmakaka}
\boldsymbol{\tilde{\sigma}}_{kk^{\prime}}=\left(            \begin{array}{cc}
               \openone_{2\times2} & -2r\,\sigma_z \\
                -2r\,\sigma_z &\openone_{2\times2}
               \end{array}
             \right)\,+\mathcal{O}(h^2),
 \end{equation}            
where $r = {}_0\beta_{kk^{\prime}}\,N\,\cos\theta $ with $\theta=2\pi\,k/(k+k^{\prime})$, and $\boldsymbol{\sigma}_z$ is the Pauli matrix. The matrix $\boldsymbol{\tilde\sigma}_{kk^{\prime}}$ corresponds to a two-mode squeezed state with squeezing parameter $r\propto N\,h$. It is possible to define an effective Hamiltonian which gives rise to such state,
\begin{align}\label{eq:ham1}
 H_{\rm int} = \mathbb{X}_{k,k'}^T\,\boldsymbol{\tilde{\sigma}}_{kk^{\prime}}\, \mathbb{X}_{k,k'} 
\end{align}
with $\mathbb{X}_{k,k'} = (Q_k,P_k,Q_{k'},P_{k'})^T$ as defined in Eq. (\ref{xp}). It is possible to generate high degrees of entanglement by increasing the number $N$ of repetitions reaching squeezing degrees as high as ($r\simeq 1/2$). However, for large $N$ second order terms in the Bogoliubov series expansion, which are proportional to $N^2\,h^2$,  can no longer be neglected. These terms introduce thermal noise in the two-mode squeezed state having negative effects in the preparation of the cluster state. However, in section \ref{sec:experiment}, we will show that a cluster state containing significant amounts of squeezing and entanglement can be generated with realistic experimental parameters.


\subsubsection*{Discrete and continuous motion}\label{subsec:continuous}
The entanglement created at first  order in $h$ between any two modes with different quantum number after any travel scenario is proportional to the first order correction to beta coefficient $B^{(1)}_{kk'}$, see Ref. \cite{nicoivy}. In the case of sharp onsets and offsets of acceleration, the coefficient $B^{(1)}_{kk'}$ is in turn proportional to the inertial-to-uniformly-accelerated first order correction $\mrbeta^{(1)}_{kk'}$:
\begin{align}
\mrbeta^{(1)}_{kk'}=\frac{\sqrt{k\,k'}}{(k+k')^3}(1-(-1)^{k+k'})\,h
\end{align}
All the above can be generalised to the case of accelerations that change continuously. In particular, if the cavity moves with a sinusoidal oscillation at a frequency $\omega_d$, then resonances appear for all the oddly separated modes that satisfy $\omega_d=\omega_k +\omega_{k'}$.
The formalism for the case of a rigid cavity that oscillates as a whole can be found in Ref.~\cite{jormadaniele}. Notice that the timescale involved in the changes of acceleration must be comparable to the characteristic timescale of the cavity, i.e. $\tau\simeq 2\,L/c$.
  
\subsubsection*{A single oscillating cavity wall}
Next, we consider the case in which only one wall of the cavity is oscillating with frequency $\omega_d=p\,\omega_1$, where $p\in\mathbb{N}$. The Bogoliubov coefficients for this case can be computed using a perturbative expansion where now 
\begin{equation}
\epsilon=\frac{\delta L}{L} \label{eq:epsilon} \ll 1
\end{equation}
is the perturbation parameter. That is, we are assuming small oscillations, as in the dynamical Casimir effect experiment \cite{casimirwilson}. Notice that in harmonic motion, the maximum acceleration is given by $a_{max}=\delta L\,\omega_d^2$. Therefore, our parameter $h_{max}$ is given by
\begin{equation}
h_{max}=\epsilon\,\frac{\omega_d^2\,L^2}{c^2} 
\end{equation}
The Bogoliubov coefficients for this case are found in Ref.~\cite{koreancasimir}. In particular, assuming that the time $T$ of the oscillation is long enough,
\begin{equation}
\omega_1\,T>>1\, , \label{eq:time}
\end{equation}
we find that the beta coefficients are given by
\begin{equation}
B^{(1)}_{kk'}=\frac{\epsilon\,\omega_1\,T}{2}\sqrt{\frac{k}{k'}}\; \delta_{k', p-k}\, ,\label{eq:betasosc}
\end{equation}
where $\delta_{k', p-k}$ is the Kronecker delta.

\subsection*{CV cluster states\label{cluster:state:section}}\noindent
Cluster states are a particular type of multipartite entangled state that, together with potentially simple measurements, are universal for quantum computation \cite{raussendorf01}. In the case of cavities, information can be encoded in the cavity field modes which are quantum continuous variables (CV) \cite{braunstein, minion1}. In this section we review the stabiliser formalism for CV cluster states and its relationship with two-mode squeezing operations.

\subsubsection*{The Stabiliser Formalism for CV cluster states}
A particularly powerful and efficient way to describe cluster states is via the stabiliser formalism \cite{gottesman97}. In this technique, a quantum state $\ket{\psi}$ is described completely by a set of $n$ operators $S_j$, called stabiliser operators, of which $\ket{\psi}$ is an eigenstate with eigenvalue $+1$:
\begin{align}
 S_j \ket{\psi} = \ket{\psi}\, , \qquad\forall j=1,\ldots, n\, .
\end{align}
The stabiliser operators $S_j$ generate an Abelian group of cardinality $2^n$. All the members of this group stabilise $\ket{\psi}$.

For continuous variables, the individual information carrying systems are quantum field modes or quantum harmonic oscillators. We can define the Heisenberg-Weyl operators
\begin{align}\label{eq:hw}
 X(q) = \exp\left( -\frac{i}{\hbar} q P \right) \quad\text{and}\quad Z(p) = \exp\left( \frac{i}{\hbar} p Q \right) ,
\end{align}
where $q,p\in\mathbb{R}$ are the continuous variables, and $Q$ and $P$ the canonical position and momentum operators of the system. The stabiliser operator for a squeezed field mode $j$ can then be written as \cite{koklovett}
\begin{align}
 S_j = X_j(\Delta) Z_j(-i\Delta^{-1})\, ,
\end{align}
where $\Delta = e^r$ is the squeezing strength, with $r$ the squeezing parameter. The single operator $S_j$ fully determines the quantum state of mode $j$. When $\Delta=1$ the field mode $j$ is in the vacuum state $\ket{0}_j$. The controlled-phase gate between two modes $j$ and $k$ is given by the operator
\begin{align}
 CZ_{jk} = \exp\left( \frac{i}{\hbar} Q_j Q_k \right) \, .
\end{align}
We can formally create continuous-variable cluster states by preparing a set of modes in momentum-squeezed states with $\Delta_j \gg 1$, and applying the controlled-phase gate $CZ_{jk}$ between all the modes that we wish to entangle. This creates new stabiliser operators 
\begin{align}
 S_j' = CZ_{jk} S_j CZ_{jk}^\dagger = X_j (\Delta) Z_k (\Delta) Z_j (-i\Delta^{-1}) \, .
\end{align}
Expressing this in exponential form using Eq.~(\ref{eq:hw}) then leads to the stabiliser condition for cluster states
\begin{align}\label{eq:ppp}
 \langle P_j - \sum_k A_{jk} Q_k\rangle \to 0\, , \quad\text{or}\quad \langle\mathbb{P} - \mathbf{A}\cdot\mathbb{Q}\rangle \to 0\, ,
\end{align}
where $\mathbb{Q} = (Q_1,\ldots,Q_M)$ and $\mathbb{P} = (P_1,\ldots,P_M)$, and $\mathbf{A}$ is the adjacency matrix for the cluster. However, ideal continuous-variable cluster states that satisfy Eq. (\ref{eq:ppp}) are unphysical, and there are various inequivalent ways to approach these states. While the generation based on the controlled-phase gates is conceptually the simplest way to generate CV cluster states, from a physical point of view the required entanglement is easier created using two-mode squeezing. We can write a two-mode squeezing operator as
\begin{equation}
\mathcal{S}_{jk} =\operatorname{exp} \big(-\xi \hat{a}^{\dagger}_j\hat{a}^{\dagger}_k +\xi^* \hat{a}_j\hat{a}_k\big).
\end{equation}
Using the Bloch-Messiah reduction \cite{braunstein2005}, a general multi-mode squeezing operator can then be constructed (up to local phase shifts) according to
\begin{equation}
\mathcal{S} = \operatorname{exp}(-i\mathcal{H}t),
\end{equation}
where $t$ is the interaction time and $\mathcal{H}$ is a bilinear interaction Hamiltonian
\begin{equation}\label{eq:ham2}
H = \sum_{jk} \left[\hat{a}_j B_{jk} \hat{a}_k + \hat{a}^{\dagger}_j B^*_{jk} \hat{a}^{\dagger}_k\right]. 
\end{equation}
Such a Hamiltonian leads to a complex adjacency matrix $\mathcal{Z}$ for a graph state \cite{menicucci11} 
\begin{equation}\label{eq:zet}
\mathcal{Z} =i\,e^{-2B} \simeq i\,\openone - 2\,i\, B^{(1)} +\mathcal{O} (h^2),
\end{equation}
where we used that in our case $h << 1$. The matrix $\mathcal{Z}$ defines an undirected graph called an H-graph that approaches a CV cluster state up to local phase shifts in the mathematical limit $r\rightarrow\infty$. In particular a succession of commuting two-mode squeezing operations of equal squeezing parameter generates an H-graph state which is bipartite, that is, all the nodes in the graph can be split into two sets such that the two-mode squeezers only relate the elements of one set with the elements of the other. By introducing $\pi/2$ phase shifts on the elements of one set, the H-graph state is transformed into a continuous variable cluster state\cite{minion3}. As we will see in detail in the Results section, in the superconducting circuit scenario that we consider in this paper the phase shift is implemented by a rotation of the same pump that generates the two-mode squeezing. For the sake of simplicity, in the remainder of the paper we will focus on how to generate the H-graph state itself, assuming that appropriate rotations of the pump can always be implemented. We will assume that all the two-mode squeezing operations commute, which is an exact result up to the first order in h. Recalling Eqs. (\ref{eq:sigmakaka}) and (\ref{eq:ham1}) and comparing them with Eqs. (\ref{eq:ham2}) and (\ref{eq:zet}), we find that a succession of commuting two-mode squeezing operations generates a matrix $B^{(1)}$ which is proportional to the first order terms of the Bogoliubov coefficients. The role of the second order deviations from the ideal scheme will be discussed in detail in the Results section. Cluster states can be diagrammatically represented by a graph in which each node corresponds to a party and the edges between nodes represent entanglement.
 
\section*{Results}\noindent
In this section we show how to produce a CV Gaussian cluster state through motion of the cavity.  In principle, arbitrarily large grids of entangled modes can be achieved.  We will also discuss possible limitations.
\subsubsection*{Step one: establishing bipartite links}
We start by choosing an \textit{odd} prime number $p\gg1$. As we will discuss later, this choice allows for a more symmetric definition of the state. We then choose two modes $k,k'$ that satisfy $p=k+k^{\prime}$.
We then notice that the first order correction to the Bogoliubov beta coefficient $B^{(1)}_{kk'}$ of an arbitrary trajectory  is proportional to 
\begin{align}
B^{(1)}_{kk'}\propto& \frac{\sqrt{kk'}}{(k+k')^3}h=\frac{\sqrt{k(p-k)}}{p^3}h\label{beta:coefficient:basic}
\end{align}
Once $p$ is fixed, we see that $B^{(1)}_{kk'}$ achieves a maximum for the couple of modes $k=\frac12 ({p-1})$ and $k'=\frac12({p+1})$ and takes the value $B^{\text{Max}}_{kk'}\sim{(2p^2)^{-1}}$.

In the following, we list the operations needed to create the multipartite entangled state i.e., we describe the motion of the boundaries in terms of total time of travel scenario. The cavity undergoes an arbitrary travel scenario that lasts a total proper time
\begin{align}
T_1=\frac{2\,L}{c\,p}\, .
\end{align}
As found in Ref.~\cite{relgates} and explained above, all pairs of modes $k,k'$ that satisfy $k+k'=p$ will have an entanglement resonance. At this early stage of the protocol we must notice that, unfortunately, the same happens also for all the couples of modes $l,l'$ which satisfy $l+l'=qp$ where $q\in\mathbb{N}$. In fact, for any $q$ there are combinations of $l$ and $l'$ for which we obtain
\begin{align}
T_q=\frac{2\pi q}{\omega_l+\omega_{l'}}=\frac{2\,L}{c\,p}\, ,
\end{align}
and therefore, all these extra couples of modes also undergo an entanglement resonance. Note that $q$ must be an odd number since $pq$ must be odd and thus, no entanglement is generated to first order in $h$.  We notice though that for these extra modes 
\begin{align}
B^{(1)}_{ll'}\propto& \frac{\sqrt{l(qp-l)}}{q^3p^3}h,
\end{align}
which in general means that, already for $q=3$, the coefficient $B^{(1)}_{ll'}$ is approximatively $3\%$ of the value of $B^{\text{Max}}_{kk'}$ (which has $q=1$). 

\begin{figure}[h]
\includegraphics[width=\columnwidth]{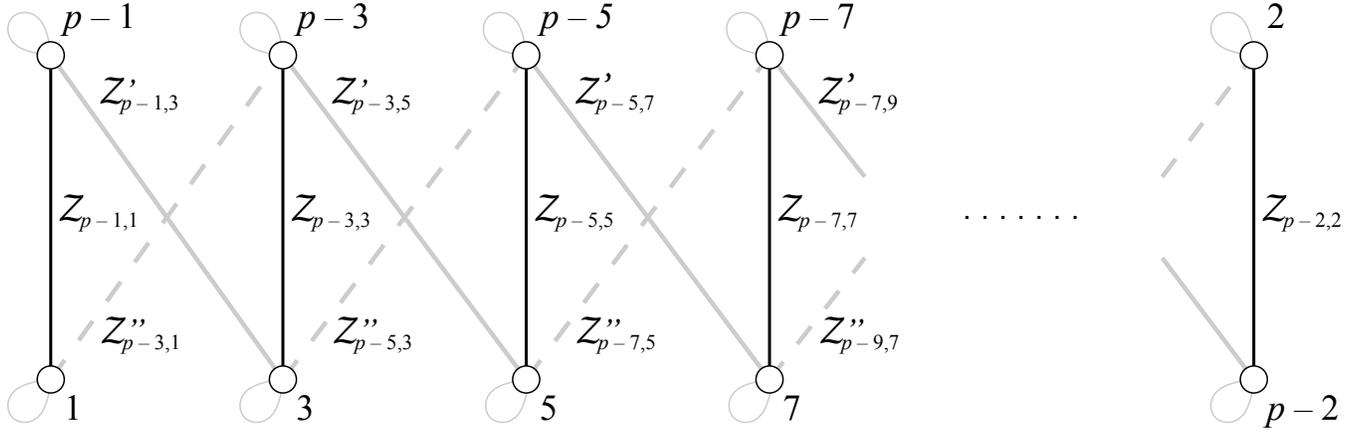}
\caption{The CV H-graph state generated by the travel scenario described in the text. The vertical edges are created by $T_1$, the solid grey diagonal edges by  $T_1'$, and the dashed edges by $T_1''$. This creates a ladder graph.}
\label{fig:graph}
\end{figure}

We can repeat the travel scenario any number of times in order to linearly increase the entanglement between the modes. This allows us to engineer entanglement between the following modes:
\begin{eqnarray}
\text{\framebox{$k=1$}} & \overset{\text{ent.}}{\longleftrightarrow} & \text{\framebox{$k'=p-1$}}\nonumber\\
\text{\framebox{$k=2$}} &\longleftrightarrow & \text{\framebox{$k'=p-2$}}\nonumber\\
\,\,\,\,\,\,\,\,\,\,\,\,\vdots & & \,\,\,\,\,\,\,\,\,\,\,\,\vdots \nonumber\\
\text{\framebox{$k=\frac{p-1}{2}$}} & \longleftrightarrow & \text{\framebox{$k'=\frac{p+1}{2}$}}. \label{first:shake}
\end{eqnarray}
Here, all double arrows represent the (``large'') entanglement created between two modes. This entanglement is shown in Fig.~\ref{fig:graph} as vertical black edges.

\subsubsection*{Step two: creating a chain of entangled modes}
We next proceed to increase the entanglement links in \eqref{first:shake},
and let the cavity undergo a new travel scenario with total proper time
\begin{align}
T^{\prime}_1=\frac{2\,L}{c\,(p-2)}\, ,
\end{align}
which will entangle all couples of modes $k,k^{\prime}$ which satisfy $k+k^{\prime}=p-2$. We can again repeat the travel scenario any number of times to achieve appreciable entanglement between the modes
\begin{eqnarray}
\text{\framebox{$k=1$}} & \longleftrightarrow & \text{\framebox{$k'=p-1$}}\nonumber\\
\updownarrow & & \nonumber\\
\text{\framebox{$k'=p-3$}} & \longleftrightarrow & \text{\framebox{$k=3$}}\nonumber\\
 & & \updownarrow\nonumber\\
\text{\framebox{$k=5$}} & \longleftrightarrow & \text{\framebox{$k'=p-5$}}\nonumber\\
\updownarrow & & \nonumber\\
\text{\framebox{$k'=p-7$}} & \longleftrightarrow & \text{\framebox{$k=7$}}\nonumber\\
 & & \updownarrow\nonumber\\
\text{\framebox{$k=9$}} & \longleftrightarrow & \text{\framebox{$k'=p-9$}}\nonumber\\
\vdots& &  \vdots
\end{eqnarray}
Again, the double arrows indicate entanglement, and is shown in Fig.~\ref{fig:graph} as diagonal dashed grey edges.

\subsubsection*{Step three: creating ladders of entangled modes}
Following the reasoning above, we now choose a final travel scenario with total proper time
\begin{align}
T^{\prime\prime}_1=\frac{2\,L}{c (p+2)}.
\end{align}
Once the travel scenario is repeated the necessary amount of times one achieves the multipartite state of the form
\begin{eqnarray}
\text{\framebox{$k=1$}} & \longleftrightarrow & \text{\framebox{$k'=p-1$}}\nonumber\\
\updownarrow & & \updownarrow\nonumber\\
\text{\framebox{$k'=p-3$}} & \longleftrightarrow & \text{\framebox{$k=3$}}\nonumber\\
\updownarrow & & \updownarrow\nonumber\\
\text{\framebox{$k=5$}} & \longleftrightarrow & \text{\framebox{$k'=p-5$}}\nonumber\\
\updownarrow & & \updownarrow\nonumber\\
\text{\framebox{$k'=p-7$}} & \longleftrightarrow & \text{\framebox{$k=7$}}\nonumber\\
\updownarrow & & \updownarrow\nonumber\\
\text{\framebox{$k=9$}} & \longleftrightarrow &\text{\framebox{$k'=p-9$}}\nonumber\\
\vdots&  &\vdots \nonumber\\
\updownarrow & & \updownarrow\nonumber\\
\text{\framebox{$k=p-2$}} & \longleftrightarrow & \text{\framebox{$k'=2$}}\label{basic:structure}
\end{eqnarray}
This entanglement is shown in Fig.~\ref{fig:graph} as diagonal solid grey edges. The resulting H-graph has a ladder configuration. The mode $p$ itself does not participate in the graph state.

\subsubsection*{Selecting contributing entangled modes}
In the previous section we have achieved a multipartite entangled state which has the ladder structure shown in Eq.~\eqref{basic:structure} and Fig.~\ref{fig:graph}.
We have already noted that the entanglement between pairs of modes in Eq.~\eqref{basic:structure} is not the same as quantified by $B^{(1)}_{kk^{\prime}}$. We would like to have a state where each entangled pair of modes exhibits the same amount of entanglement. 

Since the aim is to obtain a multipartite state with entanglement between modes that forms a particular structure, we need to guarantee that the entanglement between the modes in the structure is much higher than the entanglement with other modes. For this reason, we have chosen a large prime number $p\gg1$, and we keep all pairs of modes $k,k'$ that satisfy
\begin{eqnarray} 
\frac{|B^{(1)}_{kk'}-B^{(1)}_{\text{Max}}|}{B^{(1)}_{\text{Max}}}\geq \Lambda \label{bound:on:b},
\end{eqnarray}
where $\Lambda$ sets a limit on the minimum amount of entanglement which we want to generate. The value of $\Lambda$ sets the minimum value of $k$ -denoted by $k_{\lambda_{p}}$- that we consider and the corresponding number $\lambda_p$ of mode pairs, which is given by:
\begin{eqnarray}
 \lambda_{p}=\left|\frac{p-k_{\lambda_{p}}}{2}\right|. \label{bound:on:k}
\end{eqnarray}
Notice that every time the cavity undergoes some travel scenario, $\lambda_{p}$ will depend on the prime $p$ and will be different for different sets of modes. In addition, we have chosen numbers $p-2,p,p+2$. Clearly, for $p\gg1$ the bounds $\lambda_{p},\lambda_{p-2},\lambda_{p+2}$ will be identical given a choice of $\Lambda$. We can therefore rewrite \eqref{basic:structure} as
\begin{eqnarray}
\text{\framebox{$k=\frac{p-1}{2}$}}& \longleftrightarrow & \text{\framebox{$k'=\frac{p+1}{2}$}}\nonumber\\
\updownarrow  \,\,\,\,\,\, \,\,\,\,\,\, \,\,\,\,\,\,& & \,\,\,\,\,\, \,\,\,\,\,\, \,\,\,\,\,\,\updownarrow\nonumber\\
\text{\framebox{$k'=\frac{p+5}{2}$}} & \longleftrightarrow & \text{\framebox{$k=\frac{p-5}{2}$}} \nonumber\\
\updownarrow  \,\,\,\,\,\, \,\,\,\,\,\, \,\,\,\,\,\,& & \,\,\,\,\,\, \,\,\,\,\,\, \,\,\,\,\,\, \updownarrow\nonumber\\
\text{\framebox{$k=\frac{p-9}{2}$}} & \longleftrightarrow & \text{\framebox{$k'=\frac{p+9}{2}$}} \nonumber\\
\updownarrow  \,\,\,\,\,\, \,\,\,\,\,\, \,\,\,\,\,\,& &  \,\,\,\,\,\, \,\,\,\,\,\, \,\,\,\,\,\,\updownarrow\nonumber\\
\text{\framebox{$k'=\frac{p+13}{2}$}} & \longleftrightarrow & \text{\framebox{$k=\frac{p-13}{2}$}} \nonumber\\
\updownarrow  \,\,\,\,\,\, \,\,\,\,\,\, \,\,\,\,\,\,& \ldots & \,\,\,\,\,\, \,\,\,\,\,\, \,\,\,\,\,\, \updownarrow\nonumber\\
\text{\framebox{$k=k_{\lambda_p}$}} & \longleftrightarrow & \text{\framebox{$k'=k'_{\lambda_p}$}}.\label{truncated:basic:structure}
\end{eqnarray}
If $\lambda_p$ is chosen tight enough, all modes in \label{truncated:basic:structure} will share almost the same amount of entanglement.
\subsection*{Step four: ``Propagating'' the entanglement}
In this subsection we show how to propagate the entanglement from the basic structure we have built in Eq.~\eqref{basic:structure} and Fig.~\ref{fig:graph}, to build a regular structure.

\subsubsection*{Enlarging the structure of the multipartite state}
We have shown how we can create a H-graph state in which the  modes are entangled according to a ladder topology (a ladder graph). We will now proceed to show how to extend the ladder graph to a square lattice graph. We choose another prime $p'>p$ and $p=p'-\Delta'$.  Since $p'\gg p$ would allow for substantially different amounts of entanglement, we choose a prime $p'$ of the same order of magnitude as $p$. Among the modes that will be entangled in the ladder graph generated by $p$ there will be a mode $k=\frac12{(p+a)}$ that coincides with mode of $k=\frac12 ({p'-b})$ in the ladder graph generated by $p'$ (with $a,b\in\mathbb{N}$). If the mode $k=\frac12(p+a)=\frac12({p'-b})$ are within the ranges defined by $\lambda_{p}$ and $\lambda_{p'}$, then we can use the travel scenario. with total proper time
\begin{align}
T^{\prime\prime\prime}=\frac{2\,L}{c\,p'}\, .
\end{align}
Without loss of generality, we can assume that ${p+1}={p'-(\Delta'-1)}$. Therefore
{\tiny
\begin{align}
& &\text{\framebox{$k=\frac{p-1}{2}$}} &\,\,\,\,\,\,\,\,\,\,\,\,\,\,\,\,\,\, \longleftrightarrow & \text{\framebox{$k'=\frac{p+1}{2}=\frac{p'-(\Delta'-1)}{2}$}} & \,\,\,\,\,\,\,\,\,\,\,\,\,\,\,\,\,\,\longleftrightarrow & \text{\framebox{$\frac{p'+(\Delta'-1)}{2}$}} \nonumber\\
& &\updownarrow  \,\,\,\,\,\, \,\,\,\,\,\, \,\,\,\,\,\,& & \updownarrow\,\,\,\,\,\, \,\,\,\,\,\, \,\,\,\,\,\, \,\,\,\,\,\,& &\nonumber\\
\text{\framebox{$\frac{p'+(\Delta'-5)}{2}$}} &\,\,\,\,\,\,\,\,\,\,\,\,\,\,\,\,\,\, \longleftrightarrow &\text{\framebox{$k'=\frac{p+5}{2}=\frac{p'-(\Delta'-5)}{2}$}} & \,\,\,\,\,\,\,\,\,\,\,\,\,\,\,\,\,\,\longleftrightarrow & \text{\framebox{$k=\frac{p-5}{2}$}} & & \nonumber\\
& &\updownarrow  \,\,\,\,\,\, \,\,\,\,\,\, \,\,\,\,\,\,& &  \updownarrow  \,\,\,\,\,\, \,\,\,\,\,\, \,\,\,\,\,\, \,\,\,\,\,\,& &\nonumber\\
& &\text{\framebox{$k=\frac{p-9}{2}$}} & \,\,\,\,\,\,\,\,\,\,\,\,\,\,\,\,\,\,\longleftrightarrow & \text{\framebox{$k'=\frac{p+9}{2}=\frac{p'-(\Delta'-9)}{2}$}} & \,\,\,\,\,\,\,\,\,\,\,\,\,\,\,\,\,\,\longleftrightarrow & \text{\framebox{$\frac{p'+(\Delta'-9)}{2}$}}\nonumber\\
& &\updownarrow  \,\,\,\,\,\, \,\,\,\,\,\, \,\,\,\,\,\,& &  \updownarrow  \,\,\,\,\,\, \,\,\,\,\,\, \,\,\,\,\,\, \,\,\,\,\,\,& &\nonumber\\
\text{\framebox{$\frac{p'-(\Delta'-13)}{2}$}} & \,\,\,\,\,\,\,\,\,\,\,\,\,\,\,\,\,\,\longleftrightarrow & \text{\framebox{$k'=\frac{p+13}{2}=\frac{p'+(\Delta'-13)}{2}$}} & \,\,\,\,\,\,\,\,\,\,\,\,\,\,\,\,\,\,\longleftrightarrow & \text{\framebox{$k=\frac{p-13}{2}$}} & & \nonumber\\
& &\updownarrow  \,\,\,\,\,\, \,\,\,\,\,\, \,\,\,\,\,\,& \ldots &  \updownarrow \,\,\,\,\,\, \,\,\,\,\,\, \,\,\,\,\,\, \,\,\,\,\,\,& &\label{intermidiate:structure}
\end{align}
}%
where we do not specify the endpoint of the chain because it now depends on $\lambda_p$ and $\lambda_{p'}$. We now choose another prime $p''>p$ and $p=p''+\Delta''$ and repeat the above process to get the  structure
{\tiny
\begin{align}
\text{\framebox{$\frac{p''-(\Delta''-1)}{2}$}} & \,\,\,\,\,\,\,\,\,\,\,\,\,\,\,\,\,\,\longleftrightarrow &\text{\framebox{$k=\frac{p-1}{2}=\frac{p''+(\Delta''-1)}{2}$}} & \,\,\,\,\,\,\,\,\,\,\,\,\,\,\,\,\,\,\longleftrightarrow & \text{\framebox{$k'=\frac{p+1}{2}=\frac{p'-(\Delta'-1)}{2}$}} & \,\,\,\,\,\,\,\,\,\,\,\,\,\,\,\,\,\,\longleftrightarrow & \text{\framebox{$\frac{p'+(\Delta'-1)}{2}$}} \nonumber\\
& &\updownarrow  \,\,\,\,\,\, \,\,\,\,\,\, \,\,\,\,\,\, \,\,\,\,\,\,& & \updownarrow \,\,\,\,\,\, \,\,\,\,\,\, \,\,\,\,\,\, \,\,\,\,\,\, \,\,\,\,\,\, & &\nonumber\\
\text{\framebox{$\frac{p'+(\Delta'-5)}{2}$}} & \,\,\,\,\,\,\,\,\,\,\,\,\,\,\,\,\,\,\longleftrightarrow &\text{\framebox{$k'=\frac{p+5}{2}=\frac{p'-(\Delta'-5)}{2}$}} & \,\,\,\,\,\,\,\,\,\,\,\,\,\,\,\,\,\,\longleftrightarrow & \text{\framebox{$k=\frac{p-5}{2}=\frac{p''+(\Delta''-5)}{2}$}} & \,\,\,\,\,\,\,\,\,\,\,\,\,\,\,\,\,\,\longleftrightarrow & \text{\framebox{$\frac{p''-(\Delta''-5)}{2}$}} \nonumber\\
& &\updownarrow  \,\,\,\,\,\, \,\,\,\,\,\, \,\,\,\,\,\, \,\,\,\,\,\,& & \updownarrow \,\,\,\,\,\, \,\,\,\,\,\, \,\,\,\,\,\, \,\,\,\,\,\, \,\,\,\,\,\, & &\nonumber\\
\text{\framebox{$\frac{p''-(\Delta''-9)}{2}$}} & \,\,\,\,\,\,\,\,\,\,\,\,\,\,\,\,\,\,\longleftrightarrow &\text{\framebox{$k=\frac{p-9}{2}=\frac{p''+(\Delta''-9)}{2}$}} & \,\,\,\,\,\,\,\,\,\,\,\,\,\,\,\,\,\,\longleftrightarrow & \text{\framebox{$k'=\frac{p+9}{2}=\frac{p'-(\Delta'-9)}{2}$}} & \,\,\,\,\,\,\,\,\,\,\,\,\,\,\,\,\,\,\longleftrightarrow & \text{\framebox{$\frac{p'+(\Delta'-9)}{2}$}}\nonumber\\
& &\updownarrow  \,\,\,\,\,\, \,\,\,\,\,\, \,\,\,\,\,\, \,\,\,\,\,\,& &  \updownarrow \,\,\,\,\,\, \,\,\,\,\,\, \,\,\,\,\,\, \,\,\,\,\,\, \,\,\,\,\,\, & &\nonumber\\
\text{\framebox{$\frac{p'-(\Delta'-13)}{2}$}} & \,\,\,\,\,\,\,\,\,\,\,\,\,\,\,\,\,\,\longleftrightarrow & \text{\framebox{$k'=\frac{p+13}{2}=\frac{p'+(\Delta'-13)}{2}$}} & \,\,\,\,\,\,\,\,\,\,\,\,\,\,\,\,\,\,\longleftrightarrow & \text{\framebox{$k=\frac{p-13}{2}=\frac{p''+(\Delta''-13)}{2}$}} & \,\,\,\,\,\,\,\,\,\,\,\,\,\,\,\,\,\,\longleftrightarrow & \text{\framebox{$\frac{p''-(\Delta''-13)}{2}$}} \nonumber\\
& &\updownarrow  \,\,\,\,\,\, \,\,\,\,\,\, \,\,\,\,\,\, \,\,\,\,\,\,& \ldots & \updownarrow \,\,\,\,\,\, \,\,\,\,\,\, \,\,\,\,\,\, \,\,\,\,\,\, \,\,\,\,\,\, & & \label{intermidiate:structure:two}
\end{align}
}%
The aim will now be to close all the links and complete the lattice of entangled modes.

\subsubsection*{Final step}
\begin{figure}[h]
\includegraphics[width=\columnwidth]{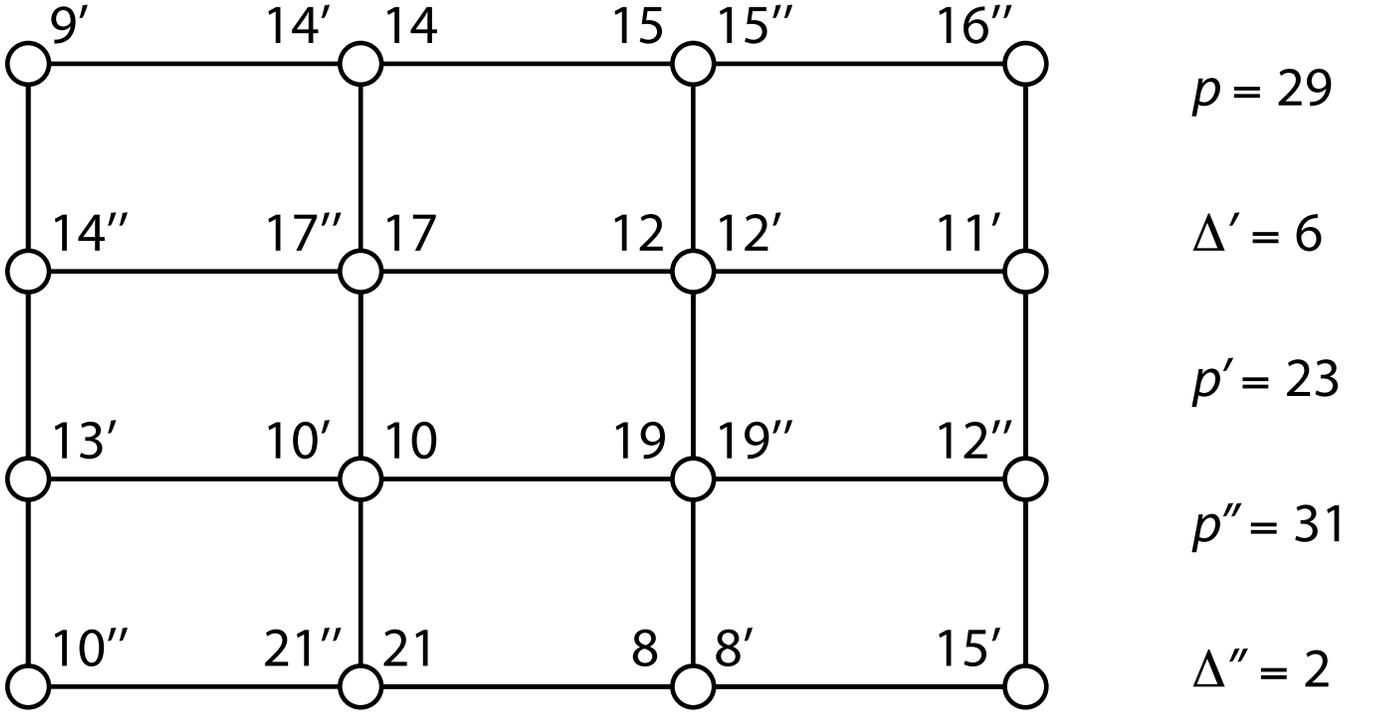}
\caption{H-graph state for $p=29$, $p'=23$, and $p''=31$, with $\Delta'=6$ and $\Delta''=2$. Each mode, indicated by an open circle, has potentially multiple labels, associated with $p$, $p'$, or $p''$.}
\label{fig:square}
\end{figure}

To complete our scheme, we notice that
\begin{align}
\frac{p''-(\Delta''-1)}{2}+\frac{p'+(\Delta'-5)}{2}&=p+\Delta'-\Delta''-2\nonumber\\
\frac{p'+(\Delta'-1)}{2}+\frac{p''-(\Delta''-5)}{2}&=p-\Delta'+\Delta''+2
\end{align}
which are both odd numbers. This again guarantees that the entanglement is created at first order in $h$.
We now let the cavity undergo two travel scenarios with proper times
\begin{align}
T^{IV}=&\frac{2\,L}{c\,(p+\Delta'-\Delta''-2)},\,\,\,\\
T^{V}=&\frac{2\,L}{c\,(p-\Delta'+\Delta''+2)}.
\end{align}
We therefore obtain our final state:
{\tiny
\begin{align}
\text{\framebox{$\frac{p''-(\Delta''-1)}{2}$}} & \,\,\,\,\,\,\,\,\,\,\,\,\,\,\,\,\,\,\longleftrightarrow &\text{\framebox{$k=\frac{p-1}{2}=\frac{p''+(\Delta''-1)}{2}$}} & \,\,\,\,\,\,\,\,\,\,\,\,\,\,\,\,\,\,\longleftrightarrow & \text{\framebox{$k'=\frac{p+1}{2}=\frac{p'-(\Delta'-1)}{2}$}} &\,\,\,\,\,\,\,\,\,\,\,\,\,\,\,\,\,\, \longleftrightarrow & \text{\framebox{$\frac{p'+(\Delta'-1)}{2}$}} \nonumber\\
\updownarrow \,\,\,\,\,\, \,\,\,\,\,\, \,\,\,\,\,\,& &\updownarrow  \,\,\,\,\,\, \,\,\,\,\,\, \,\,\,\,\,\, \,\,\,\,\,\, \,\,\,\,\,\,& & \updownarrow \,\,\,\,\,\, \,\,\,\,\,\, \,\,\,\,\,\, \,\,\,\,\,\, \,\,\,\,\,\,& & \updownarrow\,\,\,\,\,\, \,\,\,\,\,\, \,\,\,\,\,\,\nonumber\\
\text{\framebox{$\frac{p'+(\Delta'-5)}{2}$}} & \,\,\,\,\,\,\,\,\,\,\,\,\,\,\,\,\,\,\longleftrightarrow &\text{\framebox{$k'=\frac{p+5}{2}=\frac{p'-(\Delta'-5)}{2}$}} & \,\,\,\,\,\,\,\,\,\,\,\,\,\,\,\,\,\,\longleftrightarrow & \text{\framebox{$k=\frac{p-5}{2}=\frac{p''+(\Delta''-5)}{2}$}} &\,\,\,\,\,\,\,\,\,\,\,\,\,\,\,\,\,\, \longleftrightarrow & \text{\framebox{$\frac{p''-(\Delta''-5)}{2}$}} \nonumber\\
\updownarrow \,\,\,\,\,\, \,\,\,\,\,\, \,\,\,\,\,\,& &\updownarrow  \,\,\,\,\,\, \,\,\,\,\,\, \,\,\,\,\,\, \,\,\,\,\,\, \,\,\,\,\,\,& & \updownarrow \,\,\,\,\,\, \,\,\,\,\,\, \,\,\,\,\,\, \,\,\,\,\,\, \,\,\,\,\,\,& & \updownarrow\,\,\,\,\,\, \,\,\,\,\,\, \,\,\,\,\,\,\nonumber\\
\text{\framebox{$\frac{p''-(\Delta''-9)}{2}$}} & \,\,\,\,\,\,\,\,\,\,\,\,\,\,\,\,\,\,\longleftrightarrow &\text{\framebox{$k=\frac{p-9}{2}=\frac{p''+(\Delta''-9)}{2}$}} & \,\,\,\,\,\,\,\,\,\,\,\,\,\,\,\,\,\,\longleftrightarrow & \text{\framebox{$k'=\frac{p+9}{2}=\frac{p'-(\Delta'-9)}{2}$}} & \,\,\,\,\,\,\,\,\,\,\,\,\,\,\,\,\,\,\longleftrightarrow & \text{\framebox{$\frac{p'+(\Delta'-9)}{2}$}}\nonumber\\
\updownarrow \,\,\,\,\,\, \,\,\,\,\,\, \,\,\,\,\,\,& &\updownarrow  \,\,\,\,\,\, \,\,\,\,\,\, \,\,\,\,\,\, \,\,\,\,\,\, \,\,\,\,\,\,& & \updownarrow \,\,\,\,\,\, \,\,\,\,\,\, \,\,\,\,\,\, \,\,\,\,\,\, \,\,\,\,\,\,& & \updownarrow\,\,\,\,\,\, \,\,\,\,\,\, \,\,\,\,\,\,\nonumber\\\text{\framebox{$\frac{p'-(\Delta'-13)}{2}$}} & \,\,\,\,\,\,\,\,\,\,\,\,\,\,\,\,\,\,\longleftrightarrow & \text{\framebox{$k'=\frac{p+13}{2}=\frac{p'+(\Delta'-13)}{2}$}} & \,\,\,\,\,\,\,\,\,\,\,\,\,\,\,\,\,\,\longleftrightarrow & \text{\framebox{$k=\frac{p-13}{2}=\frac{p''+(\Delta''-13)}{2}$}} & \,\,\,\,\,\,\,\,\,\,\,\,\,\,\,\,\,\,\longleftrightarrow & \text{\framebox{$\frac{p''-(\Delta''-13)}{2}$}} \nonumber\\
\updownarrow \,\,\,\,\,\, \,\,\,\,\,\, \,\,\,\,\,\,& &\updownarrow  \,\,\,\,\,\, \,\,\,\,\,\, \,\,\,\,\,\, \,\,\,\,\,\, \,\,\,\,\,\,& & \updownarrow \,\,\,\,\,\, \,\,\,\,\,\, \,\,\,\,\,\, \,\,\,\,\,\, \,\,\,\,\,\,& & \updownarrow\,\,\,\,\,\, \,\,\,\,\,\, \,\,\,\,\,\,\nonumber\\
\dots \,\,\,\,\,\, \,\,\,\,\,\, \,\,\,\,\,\,& &\dots  \,\,\,\,\,\, \,\,\,\,\,\, \,\,\,\,\,\, \,\,\,\,\,\, \,\,\,\,\,\,& & \dots \,\,\,\,\,\, \,\,\,\,\,\, \,\,\,\,\,\, \,\,\,\,\,\, \,\,\,\,\,\,& & \dots\,\,\,\,\,\, \,\,\,\,\,\, \,\,\,\,\,\,\label{final:structure}
\end{align}
}%
See an example in Fig. \ref{fig:square}. So far, we have shown that, in principle, is possible to generate a large square lattice of entangled quantum field modes within a cavity. Technically, this means that we have generated an H-graph with the structure of a square. Dividing the graph in two sets such that the edges of the graph only relate elements of one set with elements of the other and applying $\pi/2$ phase shifts on the elements of one set, the complex adjacency matrix $\mathcal{Z}$ of the H-graph state -see Eq. (\ref{eq:zet})- is transformed into the adjacency matrix of a CV cluster state ~\cite{minion3}. In general, the resulting CV cluster state does not have the same form as the original H-graph, except in those cases where the adjacency matrix $A$ is self-inverse \cite{minion3,minion4}. An interesting example is a four-mode square H-graph -with one minus sign in the adjacency matrix- which is then transformed into a square CV cluster state after the phase shifts \cite{minion3,minion4}. In the next section we will show that this example is within reach of current technology in circuit QED.

\subsection*{Continuous sinusoidal motion: Experimental scenario} \label{sec:experiment}
We have developed a scheme to prepare  CV  cluster states. We now propose a simple implementation that can be achieved with current technology. The aim is to build up a quadripartite state. We choose superconducting cavities, where the boundary of the field can be controlled by external magnetic fluxes \cite{casimirtheoryv,casimircavity1,casimircavity2}.
A similar system---with only one wall---has been successfully exploited in the first experimental demonstration of the dynamical Casimir effect \cite{casimirwilson}. 

\begin{figure}[h]
\includegraphics[width=\columnwidth]{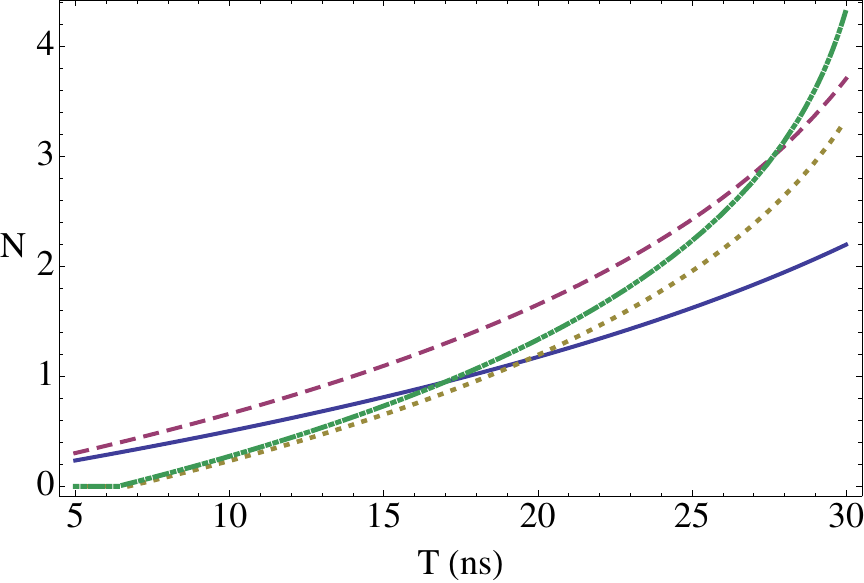}
\caption{Logarithmic negativity vs. time of the oscillatory motion for  modes $k=2$ and $k'= 5$ (solid, blue),  $k=3$ and $k'=4$ (dashed, red),  $k=2$ and $k'=3$ (dotted, yellow) and $k=4$ and $k'=5$ (dash-dotted, green). $\omega_1=2\,\pi\, 1\operatorname{GHz}$, $\epsilon=1/100$.}
\label{fig:negativity}
\end{figure}

Let us consider now a $\lambda/4$ resonator in which one of the walls is a tunable SQUID \cite{casimirtheoryv}. The spectrum is given by
\begin{equation}
\omega_k=\frac{2 \pi (k+\frac{1}{2})}{2\,L}. \label{eq:spectrum}
\end{equation}
The frequency of the fundamental mode $k=0$ is  $\omega_0=2\pi\times1\,$GHz. For the sake of convenience, we focus on a set of modes within the range of frequencies where linear amplifiers operate.
We start by driving the SQUID at $\omega_d=2\pi\times16\, \text{GHz}$. We therefore obtain
\begin{eqnarray}
\text{\framebox{$k=2$}}& \longleftrightarrow & \text{\framebox{$k'=5$}}\nonumber\\
\text{\framebox{$k=3$}} & \longleftrightarrow & \text{\framebox{$k'=4$}}.
\end{eqnarray}
We then drive the system at $\omega_d^{\prime}=12\, \text{GHz}$, entangling modes 2 and 3. Therefore
\begin{eqnarray} \label{experimental:state}
\text{\framebox{$k=2$}}& \longleftrightarrow & \text{\framebox{$k'=5$}}\nonumber\\
\updownarrow \,\,\,\,\,\, & &\nonumber\\
\text{\framebox{$k=3$}} & \longleftrightarrow & \text{\framebox{$k'=4$}} .
\end{eqnarray}
The state \eqref{experimental:state} is a multipartite entangled state. Entanglement can be extended to form a larger regular structure by means of the techniques described in this work. Within the approximations commented in \ref{subsec:continuous}, higher order entanglement resonances are completely suppressed. The time of the oscillation can be suitably chosen in order to generate a similar amount of entanglement between all the links (see Fig. \ref{fig:negativity}). In particular, the reduced state of modes $k=2$ and $k'=5$ after the first drive is given by
\begin{equation}\label{eq:statefirst}
\boldsymbol{\tilde{\sigma}}_{2\,5}=\left(            \begin{array}{cc}
               (1+(B^{(1)}_{25})^2)\openone_{2\times2} & -2\,B^{(1)}_{25}\,\sigma_z \\
                -2\,B^{(1)}_{25}\,\sigma_z &(1+(B^{(1)}_{25})^2)\openone_{2\times2}
               \end{array}
             \right)\,,
\end{equation}             
For small $h$ the state of the system remains pure to first order corresponding to a two-mode squeezed state with squeezing parameter $r_{25}=B^{(1)}_{25}$. However, when repeating the trajectory a large number of times to generate a resonance, second order terms in the expansion of the  Bogoliubov coefficients need to be included. These second order terms give rise to thermal noise quantified by $(B^{(1)}_{25})^2$ which in this case is given by Eq. (\ref{eq:betasosc}). A similar expression holds for modes $k=3$ and $k'=4$ after the first drive. The entanglement in state (\ref{eq:statefirst}) can be quantified with the logarithmic negativity which yields in this case:
\begin{equation} \label{eq:negfirst}
\mathcal{N}_{25}=\operatorname{max}[0,-\operatorname{log}(1-2\,B^{(1)}_{25}+(B^{(1)}_{25})^2)],
\end{equation}
and similarly for $\mathcal{N}_{34}$. 
Now we can drive the system at $\omega_d=2\pi\times 12\,\operatorname{GHz}$ in order to entangle modes $k=2$ and $k'=3$. However, the initial state is no longer the vacuum. After the first drive the reduced covariance matrix of modes $k=2$ and $k'=3$ is
\begin{equation}\label{eq:initialstatesecond}
\boldsymbol{\sigma}_{2\,3}=\left(            \begin{array}{cc}
               C_{25}\,\openone_{2\times2} & 0 \\
              0 & C_{34}\,\openone_{2\times2}
               \end{array}
             \right)\,,
\end{equation}
where $C_{25}=1+(B^{(1)}_{25})^2$ and $C_{34}=1+(B^{(1)}_{34})^2$.
The second drive transforms the state into
\begin{equation}\label{eq:transformedstatesecond}
\boldsymbol{\tilde{\sigma}}_{3\,4}=\left(            \begin{array}{cc}
           \boldsymbol{A} & \boldsymbol{B}\\
           \boldsymbol{B} &  \boldsymbol{A}
               \end{array}
             \right)\,,
\end{equation}
 where the matrices above have the expression 
 \begin{eqnarray}
 \boldsymbol{A}&=&\,(C_{25}+C_{34})  (1+(B^{(1)}_{23})^2)\openone_{2\times2}\nonumber\\
 \boldsymbol{B}&=&-2\,B^{(1)}_{23}[(C_{25}+C_{34})\boldsymbol{\sigma}_z +(C_{25}-C_{34})\openone_{2\times2}].
 \end{eqnarray}
Note that if the amount of squeezing and entanglement generated with the first drive were exactly the same,  i.e. $C_{25}=C_{34}=C$, then the state would be  
 \begin{equation}\label{eq:statesecond}
\boldsymbol{\tilde{\sigma}}_{2\,3}=2\,C\, \left(            \begin{array}{cc}
               (1+(B^{(1)}_{23})^2)\openone_{2\times2} & -2\,B^{(1)}_{23}\,\boldsymbol{\sigma}_z \\
                -2\,B^{(1)}_{23}\,\boldsymbol{\sigma}_z &(1+(B^{(1)}_{23})^2)\openone_{2\times2}
               \end{array}
             \right)\,.
\end{equation}             
The fact that $C_{25}\neq\,C_{34}$ adds a source of imperfection. The logarithmic negativity of the state in Eq. (\ref{eq:transformedstatesecond}) is
\begin{align}\label{eq:ent2}
\mathcal{N}_{23} =\max\left[ 0,\ln\left(\frac{2}{(C_{25}+C_{34})(1+(B^{(1)}_{23})^2)}\right)-\left((C_{25}^2+C_{34}^2)(1+10\,(B^{(1)}_{23})^2) - 2\,C_{25}\,C_{34}(1+2\,(B^{(1)}_{23})^2)\right)^{\frac12} \right].
\end{align}
Two drives suffice to build up a linear graph. A third drive at $\omega_d=20\, \operatorname{GHz}$ would entangle modes $k=4$ and $k=5$, closing the square. In this case, the entanglement would be given by a similar expression as in Eq. (\ref{eq:ent2}), replacing $ B^{(1)}_{23}$ by $ B^{(1)}_{45}$. Fig. \ref{fig:negativity} shows that a significant amount of entanglement is generated among the selected pairs of modes. The time of the drives can be chosen in such a way that all the pairs share a similar degree of entanglement.
With all the above we have generated a H-graph state with square structure. Now, if we consider a rotation of the pump by an angle $\theta$ in one of the drives, say the last one, the corresponding correlation operators are transformed to $e^{-2i\theta}a_4\,a_5$, $e^{2i\theta}a^{\dagger}_4a^{\dagger}_5$. Thus, for $\theta=\pi/2$, the corresponding two-mode squeezing operator acquires a minus sign. The effect is completely equivalent to replace  $B^{(1)}_{45}$ by $-B^{(1)}_{45}$. Therefore, the effective Hamiltonian acquires the form of the adjacency matrix of a square cluster state \cite{minion3}.
A square cluster state is a universal resource for measurement-based quantum computation, provided that homodyne detection and photon-counting are possible. Both techniques are within reach of current circuit QED technology \cite{cqed1,cqed2}.

\subsection*{Discussion}\label{sec:conclusions}
We have shown that it is possible to generate continuous-variable cluster states on electromagnetic cavity modes by choosing a suitable relativistic motion of the cavity. The entanglement  grows linearly in time. The size of the lattice is determined by the choice of initial driving frequency, the amount of entanglement required between every mode and the desired length-to-height ratio of the square cluster. As a first experimental implementation, we propose a simple example of a four-mode square cluster state in a superconducting resonator with tuneable boundary conditions. This scheme is within reach of current circuit QED technology and would be the first demonstration of a multipartite continuous variable cluster state in cQED. An interesting avenue of research would be the extension to regimes of high squeezing levels, such as the ones required for fault-tolerant quantum computing with CV cluster states \cite{menicucci14}.
In brief, our main contribution is to implement cluster states in relativistic quantum field theory, paving the way to relativistic quantum computing schemes. This is a step beyond the various proposed nonrelativistic implementations of continuous variable cluster states.

\section*{Acknowledgements}
The authors want to acknowledge G. Adesso, N. Friis and A. Lee for useful discussions. D.E.B. was in part supported by the UK Engineering and Physical Science Research Council grant number EP/J005762/1. I. F. and C.S acknowledge support from EPSRC (CAF Grant No. EP/G00496X/2 to I. F.). G.J. and P.D. would like to acknowledge funding from the Swedish Research Council and from the EU through the ERC and the projects PROMISCE and SQALEQIT. Financial support by Fundaci{\'o}n General CSIC (Programa ComFuturo) is acknowledged by C.S.

%

\end{document}